# THE HYDROGEN ATOM -- WAVE MECHANICS BEYOND SCHROEDINGER; ORBITALS AS ALGEBRAIC FORMULAE DERIVED IN ALL FOUR COORDINATE SYSTEMS

*J. F. Ogilvie*[*]

Centre for Experimental and Constructive Mathematics, Department of Mathematics, Simon Fraser University, Burnaby, British Columbia V5A 1S6 Canada
Escuela de Química, Universidad de Costa Rica, Ciudad Universitaria Rodrigo Facio, San Pedro de Montes de Oca, San José, 11501-2060 Costa Rica

**Abstract**

Chemists are aware of the solution of Schroedinger's equations for the hydrogen atom in only spherical polar coordinates, but the spatial variables are separable also in three other systems -- paraboloidal, ellipsoidal and spheroconical; we report here explicit algebraic solutions directly derived in ellipsoidal and spheroconical coordinates for the first time. Our solutions progress from those previously known in spherical polar but not entirely understood, through those little known in paraboloidal, to those in systems of ellipsoidal and spheroconical coordinates unknown before the present work. Applications of these solutions include angular momenta, a quantitative calculation of the discrete absorption spectrum and accurate plots of surfaces of amplitude functions.  The shape of a surface of a particular amplitude function, and even the quantum numbers in a particular set to specify such an individual function, depend on a particular chosen system of coordinates, and are therefore artefacts of that coordinate representation within wave mechanics; a choice of a coordinate system to discuss atomic or molecular properties based on the shapes of amplitude functions or their respective quantum numbers is hence arbitrary

**Resumen**

Los químicos están conscientes de la resolución de las ecuaciones de Schroedinger para el átomo de hidrógeno usando coordenadas polar esféricas. Sin embargo, las variables espaciales se pueden separar en otros tres sistemas también: coordenadas paraboloides, coordenadas elipsoides y coordenadas esferocónicas. Se reportan por primera vez las resoluciones explícitas algebraicas deducidas directamente en las coordenadas elipsoidales y esferocónicas. Las resoluciones parten de las coordenadas polar esféricas, aunque conocidas pero no completamente entendidas, mediante las poco conocidas coordenadas paraboloides, hasta los sistemas desconocidos de coordenadas elipsoidales y esferocónicas. Las aplicaciones de estas resoluciones incluyen los momentos angulares, el cálculo cuantitativo del espectro de absorción discreto y gráficos exactos de las superficies de las funciones de amplitud. La forma de la superficie de una función de amplitud particular, e incluso los números cuánticos en un conjunto particular para expresar tal función individual, dependen de un sistema seleccionado de coordenadas y por lo tanto, son artefactos de esa representación de la coordenada dentro de mecánica de ondas. Una selección de un sistema de coordenadas para discutir las propiedades atómicas o moleculares basadas en las formas de las funciones de amplitud o los números cuánticos respectivos es por tanto arbitraria.

---

[*] Corresponding author: ogilvie@cecm.sfu.ca

**key words:** hydrogen atom, wave mechanics, spherical polar coordinates, paraboloidal coordinates, ellipsoidal coordinates, spheroconical coordinates

**Palabras clave:** átomo de hidrógeno, mecánica de ondas, coordenadas polar esféricas, coordenadas paraboloides, coordenadas elipsoides, coordenadas esferocónicas

I. INTRODUCTION

The hydrogen atom is a fundamental topic not only as a solvable quantum-mechanical system in physics but especially as a basis of chemical theories about the binding within molecules and materials, because this simple system of a single atomic nucleus and its single associated electron is amenable to *exact* treatments within many frames of calculation. Advanced general mathematical software that enables not merely *computer algebra* but includes sophisticated methods to solve partial-differential equations and to form illuminating plots of the geometrical aspects of any calculable object can be brought to bear on such a mathematically well defined system to produce astonishingly diverse and profound results that expand and enlighten our understanding of the constitution of chemical matter. In this report we combine an application of symbolic computation to atomic hydrogen with a comprehensive investigation of all possible systems of coordinates so as to generate directly, for the first time, explicit solutions to Schroedinger's equations that have served as the foundation of an understanding of the nature of the chemical bond.

Distinct from quantum physics, which implies physical experiments on systems on an atomic scale, and from quantum chemistry, which generally implies a programmed calculation of electronic structure of molecules or materials with more or less fixed relative positions of atomic nuclei, quantum mechanics is recognized to imply a collection of methods of calculation, or algorithms, applicable to systems on an atomic scale [1, 2]. Although there exist at least twelve such distinct methods [3, 4], including Dirac's relativistic wave mechanics and quaternionic quantum mechanics, the methods most commonly applied within physics and chemistry are wave mechanics and matrix mechanics.

Quantum mechanics originated with the work of Heisenberg, who recognized that the fundamentally observable properties of an atomic system are the frequencies and intensities of its spectral lines; he developed matrix mechanics with the intention to avoid recourse to quantities unobservable, such as the orbits in Bohr's theory. Pauli applied a symbolic method to derive the energies of the hydrogen atom, but was unable to cope with the intensities of the discrete transitions [5]. Among two of his four seminal papers in 1926 under the title translated as *Quantisation as a Problem of Proper Values* [6] in which he introduced wave mechanics, Schroedinger solved the hydrogen atom first to obtain the energies of the discrete states, and subsequently to calculate the intensities of transitions between those states. In the first article Schroedinger derived amplitude functions in spherical polar coordinates that are known universally in physics and chemistry [6], and obtained the energies of the discrete states. In the third article [6], apart from developing a perturbation theory and calculating the intensities of transitions, Schroedinger treated that atom in paraboloidal coordinates. Although some textbooks of physics outline the latter derivation, it is entirely absent from textbooks of chemistry that, in contrast, typically discuss, at considerable length and with many explicit formulae and (generally inaccurate) figures, the solution of Schroedinger's temporally independent equation in spherical polar coordinates -- but seldom in SI units (recommended by IUPAC and IUPAP), as we accomplish here. For coordinates in two additional systems as variables in which Schroedinger's partial-differential equations are separable [7] and for which we here generate direct algebraic expressions for the first time, the solutions are investigated little and indirectly in ellipsoidal coordinates [8], and even less, without explicit formulae, in spheroconical coordinates [9]. We present solutions to one or other Schroedinger equation in all four systems in a similar form to facilitate comparison and to enable a profound understanding of the mathematics underlying any chemical applications.

Within wave mechanics a treatment of the hydrogen atom in coordinates in any system must reproduce the energies of the discrete states as being proportional to the inverse square of an integer, generally denoted $n$, thus defining an energy quantum number. The latter result is a conclusion purely from experiment, specifically the deductions made by Balmer and Rydberg from the wave lengths of spectral lines of the hydrogen atom emitted in the visible region; these spectral lines, measurable as circular frequency $\nu$ or wave length $\lambda$ in the optical spectrum and associated with transitions between states of the hydrogen atom, were fitted to a formula equivalent to

$$\Delta E = E_2 - E_1 = R\,h\,c\left(\frac{1}{n_1^2} - \frac{1}{n_2^2}\right) = h\,\nu = h\,c/\lambda$$

containing, with Planck constant $h$ and speed of light $c$, rydberg constant $R$ in wavenumber unit that (as $R_\infty$) is the most accurately known fundamental physical constant; an experimentally measured wave length $\lambda$ of photons is thus related to an energy difference $\Delta E$ between atomic states characterized with positive integers $n_1$ and $n_2$, with $n_1 < n_2$. The energies the hydrogen atom in its discrete states are hence implied to be expressible as $E = -R\,h\,c\,/\,n^2 + C$, in which $C$ is a constant that includes all other energy of the atomic system, including mass energy, that is not involved perceptibly in a transition between the states that yield an observed spectral line and that can hence be ignored for the present purposes. Without $C$, the energies of discrete states are negative because work must be done to remove an electron from a region near the atomic nucleus. We accordingly view $n$ as an *integer quantity that is purely experimentally derived*, bereft of any intrinsic theoretical significance, but which any acceptable theoretical treatment must reproduce. This formula might be the first analytical result in quantum physics, and has no inherent connection to quantum mechanics that it preceded by a few decades. We must, however, expect that any succeeding derivation of a solution of Schroedinger's equations for discrete states of the hydrogen atom in coordinates of various systems must yield parameters, parochial to each treatment, of which an appropriate combination becomes equivalent to that positive integer, $n$.

The objective of the present work is, for purpose of comparison, not only to derive directly, for the first time, the algebraic solutions of Schroedinger's equations in coordinates of all four systems [7] but also to shed new light on aspects of the first solution in spherical polar coordinates. The description of this system of coordinates, simply referred to as *polar* in Schroedinger's paper [6], as spherical polar is superior to merely spherical because the former term emphasizes the importance of the polar axis, in contrast with spheroconical coordinates that has also a sphere as a constant coordinate. We treat here the amplitude functions of the discrete states that produce the line spectra; the absorption continuum beyond the threshold of ionization and the associated amplitude functions in spherical polar coordinates are developed elsewhere [10]. For our purpose, we applied advanced mathematical software, *Maple*, to produce explicit and accurate algebraic formulae and quantitative graphical representations; to preclude possible error of transcription, pertinent formulae are presented directly as output from *Maple*. Here we present the algebraic aspects of the Schroedinger partial-differential equations in the temporally dependent or independent form in all four systems of coordinates in which these equations are separable and their solutions. In papers in a following set we show a diagram to define each coordinate system and several plots of the surfaces of the amplitude functions at a selected value.

According to the fundamental postulate of quantum mechanics, component $k$ of position or coordinate or displacement, $q_k$, of a particle might fail to commute with component $j$ of momentum, $p_j$, such that the commutator (with cartesian coordinates implied) is expressed as

$$[p_j, q_k] \equiv p_j q_k - q_k p_j = -i\, \delta_{jk}\, h / 2\pi,$$

in which appear $i = \sqrt{-1}$ and Kronecker's delta function; $\delta_{jk} = 1$ if $j = k$ or 0 otherwise. An application of this commutator to the differential operators pertinent to Schroedinger's formulation implies that either, with $q$ taken as an algebraic quantity according to a coordinate representation, $p$ becomes $-i\, h/2\pi\, \frac{\partial}{\partial q}$, or, with $p$ as an algebraic quantity according to a momentum representation, $q$ becomes $+i\, h/2\pi\, \frac{\partial}{\partial p}$. The particular choice of variables in either representation remains an option for the convenience of a particular calculation of an observable quantity; here we investigate the coordinate representation and coordinates in four systems.

## II.   HYDROGEN ATOM IN SPHERICAL POLAR COORDINATES

Although many authors have written much, in both journal articles and textbooks, about solutions of the general Coulomb problem according to Schroedinger's wave mechanics, appreciable misunderstanding of fundamental aspects of these solutions remains. Assuming an elimination of the motion of the centre of mass of the system, we begin here with Schroedinger's equation in its most pure and fundamental form in the coordinate representation, containing only four variables – three spatial coordinates $r$, $\theta$, $\phi$ in the spherical polar system and time, $t$ – and fundamental physical constants, so devoid of parameter other than reduced mass, $\mu = \frac{M_n m_e}{M_n + m_e}$ of the system with nuclear mass $M_n$ and electronic rest mass $m_e$, and atomic number, $Z = 1$ for H. If we express this equation in atomic units instead of SI units, even those physical constants might seem to disappear, but the retention of SI units enables a clear understanding of the dimensional nature of various quantities, and ultimately yields numerical values that are readily recognizable according to the known dimensions of an atomic system. As this system of coordinates is routinely explained in most textbooks on quantum mechanics in physics and chemistry, we here present merely the most relevant aspects for comparison with solutions in other systems of coordinates.

A differential operator requires an operand, which we call in Schroedinger's formulation an amplitude function with only spatial variables or a wave function with also a temporal variable, according to Schroedinger's usage [6]. For spherical polar coordinates and time, we thus provide as operand wave function $\Psi(r, \theta, \phi, t)$; the relation of these spatial coordinates to conventional cartesian coordinates $x,y,z$ is, according to ISO standard 80000-2:2009,

$$x = r \sin(\theta) \cos(\phi), \quad y = r \sin(\theta) \sin(\phi), \quad z = r \cos(\theta).$$

These coordinates are defined as radius $r$, which is the distance of reduced mass $\mu$ bearing electric charge $-e$ from the origin near which is located electric charge $+Z\,e$ and having domain $0 \leq r < \infty$, azimuthal angle $\theta$ of inclination subtended between polar axis $z$ and the radius vector having domain $0 \leq \theta < \pi$ rad, and equatorial angle $\phi$ subtended between the projection in plane $xy$ of the radius vector and cartesian axis *x* having domain $0 \leq \phi < 2\pi$ rad. The centre of mass of the atomic system coincides with the atomic nucleus at the origin of coordinates only in the limit of infinite mass of that nucleus. The jacobian for volume integrals of amplitude functions is $r^2 \sin(\theta)$. After separation of the motion of the centre of mass, Schroedinger's partial-differential equation for the H atom is expressed explicitly in SI units as follows.

$$\left( -\frac{\frac{\partial}{\partial r}\Psi(r,\theta,\phi,t)}{4\pi^2 r} - \frac{\cos(\theta)\left(\frac{\partial}{\partial \theta}\Psi(r,\theta,\phi,t)\right)}{8\pi^2 r^2 \sin(\theta)} - \frac{\frac{\partial^2}{\partial \theta^2}\Psi(r,\theta,\phi,t)}{8\pi^2 r^2} - \frac{\frac{\partial^2}{\partial \phi^2}\Psi(r,\theta,\phi,t)}{8\pi^2 r^2 \sin(\theta)^2} - \frac{\frac{\partial^2}{\partial r^2}\Psi(r,\theta,\phi,t)}{8\pi^2} \right) h^2/\mu - \frac{Z e^2 \Psi(r,\theta,\phi,t)}{4\pi \varepsilon_0 r} = \frac{i h \left( \frac{\partial}{\partial t}\Psi(r,\theta,\phi,t) \right)}{2\pi}$$

This equation contains only wave function $\Psi$ with its derivatives in spatial and temporal variables, no quantity other than the pertinent fundamental physical constants and parameters reduced mass $\mu$, and atomic number $Z$; the latter serves as a scaling factor for an atomic system comprising one atomic nucleus and one electron. Through the separation of variables and the subsequent solution of four ordinary-differential equations [6], the complete solution of this partial-differential equation for the bound states is expressed as this product,

$$\Psi(r, \theta, \phi, t) = R(r)\,\Theta(\theta)\,\Phi(\phi)\,T(t) = \psi(r, \theta, \phi)\,T(t)$$

as follows.

$$\Psi(r,\theta,\phi,t) = c \sqrt{\frac{Z \pi \mu e^2 k!}{\varepsilon_0 h^2 (k+2l+1)!}} \left( \frac{2\mu \pi e^2 Z}{(k+l+1) h^2 \varepsilon_0} \right)^{(l+1)} r^l$$

$$\text{LaguerreL}\left(k, 2l+1, \frac{2\pi \mu e^2 Z r}{h^2 \varepsilon_0 (k+l+1)}\right) e^{\left(-\frac{\pi \mu e^2 Z r}{h^2 \varepsilon_0 (k+l+1)}\right)} e^{(i m \phi)}$$

$$\sqrt{\frac{(2l+1)(l-|m|)!}{(l+|m|)!}}\; \text{LegendreP}(l,|m|,\cos(\theta))\, e^{\left(-\frac{i \mu Z^2 e^4 \pi t}{4 h^3 \varepsilon_0^2 (k+l+1)^2}\right)} \Big/ \big( 2(k+l+1)\sqrt{\pi}\,\big)$$

In the partial-differential equation and its solution, apart from atomic number $Z$ and reduced mass $\mu$ appear protonic charge $e$ and electric permittivity $\varepsilon_0$ of free space. In the solution of the separate ordinary-differential equation for each spatial variable $r$, $\theta$, $\phi$, a boundary condition imposes an integer value on each of three quantum numbers: the equation for R($r$) defines radial quantum number $k \geq 0$ as explained below, the equation for $\Theta(\theta)$ defines azimuthal quantum number $l \geq 0$ that has no restriction on its value relative to $k$, and the equation for $\Phi(\phi)$ defines equatorial (Schroedinger's usage [6]), or magnetic, quantum number $m$, with $-l \leq m \leq +l$, according to the indicated ranges. This radial quantum number $k$ is limited to this coordinate system within wave mechanics; its significance is that it specifies the number of radial nodes of product $\Psi^* \Psi$ – those radii at which $|\psi^2|$ becomes zero between $r = 0$ and $r \to \infty$, whereas azimuthal quantum number $l$ not only indicates the number of angular nodes between $\theta = 0$ and $\theta = \pi$ rad but also

J. F. OGILVIEpertains to the orbital angular momentum, *vide infra*. $\Psi^*$ implies the complex conjugate of $\Psi$. In this directly generated solution of the governing partial-differential equation, the exponential term containing distance $r$ directly contains a minus sign, implying an exponential decay of $\psi(r,\theta,\phi)$ as $r \to \infty$; if a plus sign appeared in that direct solution, that formula might still be acceptable, provided that a polynomial factor in $r$, here expressed as an associated Laguerre function [11], generally denoted $L_k^{2l+1}(r)$, as a product with associated Legendre functions of the first kind [11], generally denoted $P_l^m(\cos\theta)$, possesses an appropriate form [12].

The quantity multiplying $it$ in the exponential factor within T($t$), and hence within $\Psi(r, \theta, \phi, t)$, has the physical significance of an angular frequency, as Schroedinger recognised [6]; in this context we interpret that angular frequency as an energy $E$ divided by Planck constant $h$, which Schroedinger also considered; a further factor $2\pi$ to convert to a circular frequency yields this expression.

$$E = -\frac{1}{8}\frac{\mu Z^2 e^4}{h^2 \varepsilon_0^2 (k+l+1)^2}$$

In the denominator of this formula for energy appears a sum of two non-negative integer quantum numbers, $k$ and $l$, plus unity; that sum must consequently equal a positive integer, which *we associate with an energy quantum number*, denoted $n$, that was deduced from experiment by Balmer and Rydberg. The cluster of accompanying constants hence becomes equivalent to $R\,h\,c$, as described above. The zero of the energy scale is set according to the potential energy; V($r$) = 0 as $r \to \infty$ with no relative motion of proton and electron. As $n = k + l + 1$, the energy of a discrete state of H, defined with quantum numbers $k$, $l$, $m$, for which $E < 0$ and subject to no external influence, depends on values of $k$ and $l$ in the specified combination in this system of coordinates, and is thus independent of the value of equatorial quantum number $m$. For a H atom in the presence of an externally applied magnetic field, the energy of a state depends also on equatorial quantum number $m$, according to the Zeeman effect, either normal or anomalous (with account taken of the intrinsic angular momentum of the electron).

Like three following solutions in coordinates of other systems, the total solution, above, of the partial-differential equation contains coefficient $c$ (not speed of light) that equals any complex quantity of modulus unity such as the fourth roots of unity, $c = \pm 1, \pm\sqrt{-1}$, which appears because Schroedinger's temporally dependent equation is linear and homogeneous, or equally because the corresponding temporally independent equation has the form of an eigenvalue relation. One might equally express that coefficient as an exponential phase factor, $c = e^{i\alpha}$. The conventional choice $c = 1$ or $\alpha = 0$, which is arbitrary and lacks physical justification, signifies that some solutions $\psi(r,\theta,\phi)$, as amplitude functions from the temporally independent Schroedinger equation, might appear in a real form if $m = 0$, whereas most are complex because of the presence of factor $e^{im\phi}$, having real and imaginary parts for $m \neq 0$; with a mathematically viable alternative choice $c = \sqrt{-1}$, some amplitude functions would be purely imaginary but most would still be complex.

To heed another aspect of these wave functions that might be conventionally overlooked, consistent with Heisenberg's focus on the spectral properties of an atom, we calculate the intensity of a spectral transition, from the electronic ground state, that is proportional to the squared matrix element involving the electric dipolar (or other) moment as operator, which we express in two forms, taking cartesian coordinate $z = r\cos(\theta)$ as the direction of both the dipolar moment of the transition and the electric vector of an electromagnetic wave incident on the atom.

$$< e\,z > \equiv <k', l', m'|e\,z|k, l, m> \ = \ <0, l, 0|e\,r\cos(\theta)|0, 0, 0>$$

$$\langle e\,z \rangle \equiv \int_0^\infty \int_0^\pi \int_0^{2\pi} \Psi(r, \theta, \phi, t, k', l', m')^* \, e\, r \cos(\theta)\, \Psi(r, \theta, \phi, t, k, l, m)\, r^2 \sin(\theta)\, d\phi\, d\theta\, dr$$

In Dirac's *bracket* notation, the upper line defines a matrix element specifying quantum numbers *k*, *l*, *m* of amplitude functions of two particular combining states in the *bra* and *ket* quantities, one being the ground electronic state; the integral in the lower line expresses a general calculation directly in terms of the wave functions associated with any two states. We distinguish between a state – i.e. a spectrometric state – and an amplitude or wave function that is an artefact of wave mechanics, and of which its quantum numbers enable a calculation of the energy of that state. For the electronic transition to a state of least energy from the initial ground state in absorption, for which the terminal state has *bra* quantity is < 0, 1, 0|, an explicit integration according to the lower formula yields this result.

$$\langle ez \rangle = \frac{128\sqrt{2}\, h^2\, \varepsilon_0\, \mathrm{e}^{\left(-\frac{3\,i\,\mu\,Z^2\,e^4\,\pi\,t}{16\,h^3\,\varepsilon_0^2}\right)}}{243\, Z\, e\, \pi\, \mu}$$

Because temporally dependent wave functions appear in the preceding triple integral over spatial coordinates, the exponent in the result contains a function of time; the coefficient of $2\pi i t$ in the exponent,

$$\nu = \frac{3\,\mu\,Z^2\,e^4}{32\,h^3\,\varepsilon_0^2}$$

we interpret directly as circular frequency $\nu$ of a photon involved in a transition between the specified discrete states, rather than as an energy difference $\Delta E$ divided by Planck's constant; we thus contrast the interpretation of the corresponding coefficient, in the original wave functions before their subjection to integration, as an energy. The evaluation of those matrix elements and frequencies enables one to plot an absorption spectrum that is quantitatively accurate [10], within the Schroedinger formalism, corresponding to spectral lines in the Lyman series in the vacuum-ultraviolet region. The temporal factor in the matrix element above does not affect the intensity because the exponent vanishes in forming the square of the magnitude of the matrix element to produce the oscillator strength, or *f* value, as a dimensionless measure of intensity:

$$f = \frac{8}{3}\, \frac{\pi^2 m_e}{h\, e^2}\, \nu\, |<e\,z>|^2$$

With appropriately formed amplitude functions and operators for electric dipolar moment, the intensity is equally well calculated in other coordinate systems; *vide infra*.

### III. HYDROGEN ATOM IN PARABOLOIDAL COORDINATES

A treatment of the H atom in paraboloidal coordinates *u*, *v*, $\phi$ is typically only sketched in some textbooks of physics and is entirely absent from textbooks of chemistry. As those treatments differ in any case from the strategy of a direct solution in SI units practicable with *Maple*, we present here in a little detail the pertinent solution of Schroedinger's temporally independent equation; the temporal part is analogous to that for spherical polar coordinates. According to Spiegel's definition [13], the relations between cartesian and paraboloidal coordinates $u$, $v$, $\phi$ are

$$x = u\,v\,\cos(\phi),\ y = u\,v\,\sin(\phi),\ z = \frac{1}{2}(u^2 - v^2),\ r = \frac{1}{2}(u^2 + v^2)$$

with domains $0 \leq u < \infty$, $0 \leq v < \infty$ and $0 \leq \phi < 2\pi$ rad; the reduced mass is located at distance $r$ from the origin near which is located charge $+Z\,e$. Surfaces of constant $u$ and $v$ are paraboloids, of circular cross section, about axis $z$, opening toward $-z$ and $+z$, respectively; a surface of constant $\phi$ is a half-plane containing axis $z$. The distinguishing feature of these coordinates is the presence of these paraboloids, which makes a description of the coordinates as paraboloidal preferable to parabolic. The jacobian for volume integrals is $u\,v\,(u^2 + v^2)$. As the temporal component separates just as readily in other systems of coordinates as in spherical polar coordinates and has a common factor $\tau(t)$ that becomes interpreted in terms of energy quantum number $n$, we henceforth present the results for amplitude functions, thus involving only the spatial coordinates. Schroedinger's equation independent of time is accordingly expressed in paraboloidal coordinates as

$$-\frac{h^2 \left(\frac{\partial}{\partial u}\psi(u,v,\phi)\right)}{8\pi^2 \mu (u^2+v^2)\,u} - \frac{h^2 \left(\frac{\partial^2}{\partial u^2}\psi(u,v,\phi)\right)}{8\pi^2 \mu (u^2+v^2)} - \frac{h^2 \left(\frac{\partial}{\partial v}\psi(u,v,\phi)\right)}{8\pi^2 \mu (u^2+v^2)\,v} - \frac{h^2 \left(\frac{\partial^2}{\partial v^2}\psi(u,v,\phi)\right)}{8\pi^2 \mu (u^2+v^2)}$$

$$-\frac{h^2 \left(\frac{\partial^2}{\partial \phi^2}\psi(u,v,\phi)\right)}{8\pi^2 \mu (u^2+v^2)\,v^2} - \frac{h^2 \left(\frac{\partial^2}{\partial \phi^2}\psi(u,v,\phi)\right)}{8\pi^2 \mu (u^2+v^2)\,u^2} - \frac{Z\,e^2\,\psi(u,v,\phi)}{4\pi\varepsilon_0 \frac{u^2+v^2}{2}} = E\,\psi(u,v,\phi)$$

in which appears amplitude function $\psi(u,v,\phi)$ of these spatial variables. This partial-differential equation becomes solved on separation of the variables into three ordinary-differential equations with appropriate separation parameters. The angular solution for $\Phi(\phi)$ is the same as in spherical polar coordinates and hence contains equatorial quantum number $m$, namely,

$$\Phi(\phi) = \frac{e^{(i\,m\,\phi)}}{\sqrt{2\pi}}$$

After conversion from Whittaker functions in *Maple*'s direct solution of the ordinary-differential equations for U($u$) and V($v$) because Schroedinger's equation in these coordinates is a special case of Whittaker's differential equation, through Kummer functions to preserve the phase characteristics, to associated Laguerre functions [11] that both U($u$) and V($v$) contain, and setting the first arguments of these Laguerre functions to non-negative integers $n_1$ and $n_2$, respectively, the algebraic solutions for the coordinates of length type become

$$U(u) = (-1)^{|m|} N_u\, e^{\left(-1/2\,\frac{Z\,e^2\,\pi\mu\,u^2}{h^2\varepsilon_0(1+|m|+2n_1)}\right)} \left(\frac{\pi Z e^2 \mu}{h^2 \varepsilon_0 (1+|m|+2n_1)}\right)^{(1/2+1/2|m|)} u^{|m|}$$

$$\mathrm{LaguerreL}\!\left(n_1, |m|, \frac{Z\,e^2\,\pi\mu\,u^2}{h^2\varepsilon_0 (1+|m|+2n_1)}\right) \Big/ \mathrm{binomial}(1+|m|+n_1, n_1)$$

and

$$V(v) = N_v \, e^{\left(-1/2 \frac{\pi \mu Z e^2 v^2}{h^2 \varepsilon_0 (|m|+1+2n_2)}\right)} \left(\frac{\pi Z e^2 \mu}{h^2 \varepsilon_0 (|m|+1+2n_2)}\right)^{(1/2+1/2|m|)} v^{|m|}$$

$$\text{LaguerreL}\left(n_2, |m|, \frac{\pi \mu Z e^2 v^2}{h^2 \varepsilon_0 (|m|+1+2n_2)}\right) \Big/ \text{binomial}(1+|m|+n_2, n_2)$$

in which appear the indicated binomial coefficients. These two separate equations for U($u$) and V($v$) have equivalent forms; the energy must depend on both $n_1$ and $n_2$ in an equivalent manner: for this purpose we replace 2 $n_1$ or 2 $n_2$, wherever this quantity appear, by $n_1 + n_2$. Among alternative arguments, Bethe and Salpeter [14] associated part of nuclear charge $Z$ with one distance coordinate and the other part with the other distance coordinate; other authors use similarly convoluted explanations to achieve the same result, but the above argument is convenient here. Apart from these quantum numbers that are imposed to conform to the boundary conditions applicable to U($u$) and V($v$) through the associated Laguerre polynomials, and from $N_u$ and $N_v$ that are normalizing factors to be evaluated, the notation is similar to that applied for spherical polar coordinates. Including the correct total normalizing factor, for bound states the exact product $\psi(u,v,\phi) = U(u) \, V(v) \, \Phi(\phi)$ of the three solutions of the spatial ordinary-differential equations becomes expressed as this explicit amplitude function:

$$\psi(u, v, \phi) = c \, (-1)^{|m|} \sqrt{\frac{Z \pi \mu e^2}{\varepsilon_0 h^2}} \sqrt{\frac{2 n_1! n_2!}{(n_1+|m|)! (n_2+|m|)!}}$$

$$\left(\frac{\pi Z e^2 \mu}{h^2 \varepsilon_0 (|m|+n_1+n_2+1)}\right)^{(1+|m|)} (u v)^{|m|} e^{\left(-\frac{\pi Z e^2 \mu (u^2+v^2)}{2 h^2 \varepsilon_0 (|m|+n_1+n_2+1)}\right)} e^{(i m \phi)}$$

$$\text{LaguerreL}\left(n_1, |m|, \frac{\pi Z e^2 \mu u^2}{h^2 \varepsilon_0 (|m|+n_1+n_2+1)}\right)$$

$$\text{LaguerreL}\left(n_2, |m|, \frac{\pi Z e^2 \mu v^2}{h^2 \varepsilon_0 (|m|+n_1+n_2+1)}\right) \Big/ (\sqrt{2\pi} \, (|m|+n_1+n_2+1))$$

According to the coefficient of $t$ in the temporal exponent of the corresponding wave function (not shown), or the exponential term above containing $u^2 + v^2$, the energy associated with an amplitude function characterized with quantum numbers $n_1$, $n_2$, $m$ becomes expressed as

$$E = -\frac{\mu Z^2 e^4}{8 h^2 \varepsilon_0^2 (n_1+n_2+|m|+1)^2}$$

Quantum numbers $n_1$ and $n_2$ assume values of only non-negative integers; quantum number $m$ takes values of negative and positive integers and zero but without constraint of $n_1$ or $n_2$. In the denominator of this formula for energy appear in a sum two non-negative integers $n_1$ and $n_2$ and the absolute value of another integer quantum number $m$ and unity; this sum must consequently equal a positive integer, denoted $n$, which we again associate with an energy

J. F. OGILVIEquantum number from experiment, so $n = n_1 + n_2 + |m| + 1$. The energy associated with an amplitude function pertaining to a discrete state of H defined with these quantum numbers thus depends directly on all three local quantum numbers in combination, hence including equatorial quantum number *m even in the absence of an externally applied magnetic field*, but $|m|$ is precisely a lower limit of azimuthal quantum number *l* that appears directly in the solution of Schroedinger's equation in spherical polar coordinates.

## IV.   HYDROGEN ATOM IN ELLIPSOIDAL COORDINATES

In this coordinate system, electric charge $+ Z e$ is located at the origin that is also one focus of an ellipsoid; another focus is located at distance *d* along the positive *z* axis. The distance of mass µ of negative charge $-e$ from that origin is $r_1$; its distance from the other focus is $r_2$. Ellipsoidal coordinates are defined as reduced distances,

$$\xi = (r_1 + r_2)/d, \quad \eta = (r_1 - r_2)/d,$$

hence dimensionless, with equatorial angle $\phi$ as in spherical polar and paraboloidal coordinates; their domains are $1 \leq \xi < \infty$, $-1 \leq \eta \leq 1$ and $0 \leq \phi < 2\pi$ rad. A surface of constant $\xi$ is a confocal circular ellipsoid of revolution about axis *z*; a surface of constant $\eta$ is a corresponding circular hyperboloid of one sheet about axis *z*, and a surface of constant $\phi$ is a half plane containing axis *z*. The distinguishing feature of these coordinates is that ellipsoid that possesses two centres or foci, which makes a description of these coordinates as ellipsoidal preferable to prolate spheroidal, even though these ellipsoids represent special cases of spheroids with two equal semi-axes. The relations between these coordinates $\xi$, $\eta$, $\phi$ and cartesian coordinates *x*, *y*, *z* or radial coordinate *r* are hence

$$x = \frac{1}{2} d \sqrt{(\xi^2 - 1)(-\eta^2 + 1)} \cos(\phi), \; y = \frac{1}{2} d \sqrt{(\xi^2 - 1)(-\eta^2 + 1)} \sin(\phi), \; z = \frac{d(\eta\xi + 1)}{2}$$

$$r = \frac{d(\eta + \xi)}{2}$$

The jacobian for volume integrals is $(\xi^2 - \eta^2) \, d^3/8$. According to those definitions, Schroedinger's temporally independent equation becomes

$$\frac{1}{8} h^2 \left( \left( \frac{\partial^2}{\partial \xi^2} \psi(\xi, \eta, \phi) \right) \eta^2 \xi^4 - \left( \frac{\partial^2}{\partial \eta^2} \psi(\xi, \eta, \phi) \right) \eta^4 \xi^2 + 2 \left( \frac{\partial}{\partial \xi} \psi(\xi, \eta, \phi) \right) \eta^2 \xi^3 \right.$$

$$- 2 \left( \frac{\partial}{\partial \eta} \psi(\xi, \eta, \phi) \right) \eta^3 \xi^2 - 2 \left( \frac{\partial^2}{\partial \xi^2} \psi(\xi, \eta, \phi) \right) \eta^2 \xi^2 - \left( \frac{\partial^2}{\partial \xi^2} \psi(\xi, \eta, \phi) \right) \xi^4$$

$$+ \left( \frac{\partial^2}{\partial \eta^2} \psi(\xi, \eta, \phi) \right) \eta^4 + 2 \left( \frac{\partial^2}{\partial \eta^2} \psi(\xi, \eta, \phi) \right) \eta^2 \xi^2 - 2 \left( \frac{\partial}{\partial \xi} \psi(\xi, \eta, \phi) \right) \eta^2 \xi$$

$$\left. - 2 \left( \frac{\partial}{\partial \xi} \psi(\xi, \eta, \phi) \right) \xi^3 + 2 \left( \frac{\partial}{\partial \eta} \psi(\xi, \eta, \phi) \right) \eta^3 + 2 \left( \frac{\partial}{\partial \eta} \psi(\xi, \eta, \phi) \right) \eta \xi^2 \right.$$

$$+ \left( \frac{\partial^2}{\partial \xi^2} \psi(\xi, \eta, \phi) \right) \eta^2 + 2 \left( \frac{\partial^2}{\partial \xi^2} \psi(\xi, \eta, \phi) \right) \xi^2 - 2 \left( \frac{\partial^2}{\partial \eta^2} \psi(\xi, \eta, \phi) \right) \eta^2$$

$$- \left( \frac{\partial^2}{\partial \eta^2} \psi(\xi, \eta, \phi) \right) \xi^2 + \left( \frac{\partial^2}{\partial \phi^2} \psi(\xi, \eta, \phi) \right) \eta^2 - \left( \frac{\partial^2}{\partial \phi^2} \psi(\xi, \eta, \phi) \right) \xi^2$$

$$+ 2 \left( \frac{\partial}{\partial \xi} \psi(\xi, \eta, \phi) \right) \xi - 2 \left( \frac{\partial}{\partial \eta} \psi(\xi, \eta, \phi) \right) \eta - \left( \frac{\partial^2}{\partial \xi^2} \psi(\xi, \eta, \phi) \right) + \left( \frac{\partial^2}{\partial \eta^2} \psi(\xi, \eta, \phi) \right)$$

$$\Bigg) / (\pi^2 \mu (\eta^4 \xi^2 - \eta^2 \xi^4 - \eta^4 + \xi^4 + \eta^2 - \xi^2)) - \frac{1}{2} \frac{Z e^2 \psi(\xi, \eta, \phi)}{\pi \varepsilon_0 d (\eta + \xi)} = E \psi(\xi, \eta, \phi)$$

The variables become partially separated to form three ordinary-differential equations:

$$\frac{d^2}{d\phi^2} \Phi(\phi) = \_c_3 \Phi(\phi)$$

$$\frac{d^3}{d\xi^3} \Xi(\xi) = \Bigg( -2 \Xi(\xi) h^2 \left( \frac{d}{d\xi} \Xi(\xi) \right) \varepsilon_0 d - 4 \Xi(\xi) h^2 \left( \frac{d^2}{d\xi^2} \Xi(\xi) \right) \xi \varepsilon_0 d$$

$$- 16 E \Xi(\xi)^2 \pi^2 \mu \varepsilon_0 d \xi^5 - 16 E \Xi(\xi)^2 \pi^2 \mu \varepsilon_0 d \xi$$

$$+ h^2 \left( \frac{d^2}{d\xi^2} \Xi(\xi) \right) \xi^6 \varepsilon_0 d \left( \frac{d}{d\xi} \Xi(\xi) \right) - 3 h^2 \left( \frac{d^2}{d\xi^2} \Xi(\xi) \right) \xi^4 \varepsilon_0 d \left( \frac{d}{d\xi} \Xi(\xi) \right)$$

$$+ 3 h^2 \left( \frac{d^2}{d\xi^2} \Xi(\xi) \right) \xi^2 \varepsilon_0 d \left( \frac{d}{d\xi} \Xi(\xi) \right) - 4 h^2 \left( \frac{d}{d\xi} \Xi(\xi) \right)^2 \xi^3 \varepsilon_0 d$$

$$+ 2 h^2 \left( \frac{d}{d\xi} \Xi(\xi) \right)^2 \xi \varepsilon_0 d + 2 h^2 \left( \frac{d}{d\xi} \Xi(\xi) \right)^2 \xi^5 \varepsilon_0 d - h^2 \left( \frac{d^2}{d\xi^2} \Xi(\xi) \right) \varepsilon_0 d \left( \frac{d}{d\xi} \Xi(\xi) \right)$$

$$- 4 Z e^2 \Xi(\xi)^2 \pi \mu - 4 \pi \Xi(\xi)^2 Z e^2 \mu \xi^4 + 8 \pi \Xi(\xi)^2 Z e^2 \mu \xi^2$$

$$+ 32 E \Xi(\xi)^2 \pi^2 \mu \varepsilon_0 d \xi^3 + 4 \Xi(\xi) h^2 \left( \frac{d}{d\xi} \Xi(\xi) \right) \xi^2 \varepsilon_0 d$$

$$- 4 \Xi(\xi) h^2 \left( \frac{d^2}{d\xi^2} \Xi(\xi) \right) \xi^5 \varepsilon_0 d - 2 \Xi(\xi) h^2 \left( \frac{d}{d\xi} \Xi(\xi) \right) \xi^4 \varepsilon_0 d$$

$$+ 2 h^2 \_c_3 \Xi(\xi)^2 \xi \varepsilon_0 d + 8 \Xi(\xi) h^2 \left( \frac{d^2}{d\xi^2} \Xi(\xi) \right) \xi^3 \varepsilon_0 d \Bigg) / ($$

$$\Xi(\xi) h^2 \xi^6 \varepsilon_0 d - 3 h^2 \Xi(\xi) \xi^4 \varepsilon_0 d + 3 h^2 \Xi(\xi) \xi^2 \varepsilon_0 d - h^2 \Xi(\xi) \varepsilon_0 d)$$

$$\frac{d^2}{d\eta^2} H(\eta) = - \Bigg( 4 \eta^2 \xi Z e^2 \Xi(\xi) H(\eta) \pi \mu + 4 \eta^3 \pi \Xi(\xi) H(\eta) Z e^2 \mu \xi^2$$

$$+ 8 E \Xi(\xi) H(\eta) \pi^2 \mu \varepsilon_0 d \xi^4 - 8 \eta^4 E \Xi(\xi) H(\eta) \pi^2 \mu \varepsilon_0 d$$

$$+ 8 \eta^2 E \Xi(\xi) H(\eta) \pi^2 \mu \varepsilon_0 d - 4 \eta \pi \Xi(\xi) H(\eta) Z e^2 \mu \xi^2$$

$$- 8 E \Xi(\xi) H(\eta) \pi^2 \mu \varepsilon_0 d \xi^2 - 4 \eta^2 \pi \Xi(\xi) H(\eta) Z e^2 \mu \xi^3$$

$$+ 4 \pi \Xi(\xi) H(\eta) Z e^2 \mu \xi^3 - 4 \xi Z e^2 \Xi(\xi) H(\eta) \pi \mu - 4 \eta^3 Z e^2 \Xi(\xi) H(\eta) \pi \mu$$

$$+ 4 \eta Z e^2 \Xi(\xi) H(\eta) \pi \mu + 2 h^2 \Xi(\xi) \left( \frac{d}{d\eta} H(\eta) \right) \eta^3 \xi^2 \varepsilon_0 d$$

$$\begin{aligned}&-2\,h^2\left(\frac{d}{d\xi}\Xi(\xi)\right)H(\eta)\,\eta^2\,\xi^3\,\varepsilon_0\,d+2\,h^2\left(\frac{d}{d\xi}\Xi(\xi)\right)H(\eta)\,\eta^2\,\xi\,\varepsilon_0\,d\\&-2\,h^2\,\Xi(\xi)\left(\frac{d}{d\eta}H(\eta)\right)\eta\,\xi^2\,\varepsilon_0\,d+h^2\,\_c_3\,\Xi(\xi)\,H(\eta)\,\xi^2\,\varepsilon_0\,d\\&-h^2\,\_c_3\,\Xi(\xi)\,H(\eta)\,\eta^2\,\varepsilon_0\,d-h^2\left(\frac{d^2}{d\xi^2}\Xi(\xi)\right)H(\eta)\,\eta^2\,\xi^4\,\varepsilon_0\,d\\&+2\,h^2\left(\frac{d^2}{d\xi^2}\Xi(\xi)\right)H(\eta)\,\eta^2\,\xi^2\,\varepsilon_0\,d+8\,\eta^4\,E\,\Xi(\xi)\,H(\eta)\,\pi^2\,\mu\,\varepsilon_0\,d\,\xi^2\\&-8\,\eta^2\,E\,\Xi(\xi)\,H(\eta)\,\pi^2\,\mu\,\varepsilon_0\,d\,\xi^4-2\,h^2\,\Xi(\xi)\left(\frac{d}{d\eta}H(\eta)\right)\eta^3\,\varepsilon_0\,d\\&+2\,h^2\left(\frac{d}{d\xi}\Xi(\xi)\right)H(\eta)\,\xi^3\,\varepsilon_0\,d+2\,h^2\,\Xi(\xi)\left(\frac{d}{d\eta}H(\eta)\right)\eta\,\varepsilon_0\,d\\&-2\,h^2\left(\frac{d}{d\xi}\Xi(\xi)\right)H(\eta)\,\xi\,\varepsilon_0\,d+h^2\left(\frac{d^2}{d\xi^2}\Xi(\xi)\right)H(\eta)\,\xi^4\,\varepsilon_0\,d\\&-h^2\left(\frac{d^2}{d\xi^2}\Xi(\xi)\right)H(\eta)\,\eta^2\,\varepsilon_0\,d-2\,h^2\left(\frac{d^2}{d\xi^2}\Xi(\xi)\right)H(\eta)\,\xi^2\,\varepsilon_0\,d\\&+h^2\left(\frac{d^2}{d\xi^2}\Xi(\xi)\right)H(\eta)\,\varepsilon_0\,d\Big)\Big/(h^2\,\Xi(\xi)\,\varepsilon_0\,d\,(\eta^4\,\xi^2-\eta^4-2\,\eta^2\,\xi^2+2\,\eta^2+\xi^2-1\,)\\&)\end{aligned}$$

The solution of the first equation, for the angular variable, is the same as that displayed above, which enables a value for separation constant $\_c_3$ introduced automatically in *Maple*,

$$\_c_3 = -m^2$$

which we substitute into the other two equations involving variables $\xi$ and $\eta$. As the other two differential equations, that for $\Xi(\xi)$ being of second order and that for $H(\eta)$ being of third order, are coupled, we solve them simultaneously as a system to yield, after elimination of physically unacceptable solutions,

$$\Xi(\xi) = N_\xi\,(\xi^2-1)^{\left(\frac{|m|}{2}\right)} e^{\left(-\frac{Z\,d\,\xi}{2\,n\,a_0}\right)}\\ \text{HeunC}\left(\frac{2\,Z\,d}{n\,a_0},|m|,|m|,\frac{2\,Z\,d}{a_0},\frac{m^2}{2}-\frac{Z\,d}{a_0}+\_C1-\frac{Z^2\,d^2}{4\,n^2\,a_0^{\,2}},\frac{\xi}{2}+\frac{1}{2}\right)$$

and

$$H(\eta) = N_\eta\,(1-\eta^2)^{\left(\frac{|m|}{2}+\frac{1}{2}\right)} e^{\left(-\frac{Z\,d\,(\eta+1)}{2\,n\,a_0}\right)}\\ \text{HeunC}\left(\frac{2\,Z\,d}{n\,a_0},|m|,|m|,\frac{2\,Z\,d}{a_0},\frac{m^2}{2}-\frac{Z\,d}{a_0}+\_C1-\frac{Z^2\,d^2}{4\,n^2\,a_0^{\,2}},\frac{\eta}{2}+\frac{1}{2}\right)$$

In these two solutions of which the simple form is similar to that of R(*r*) in spherical polar coordinates despite their evolution from coupled differential equations of second and third order, $N_\xi$ and $N_\eta$ are normalizing factors to be evaluated; _C1 is a parameter that must satisfy an appropriate condition. These two solutions have parallel forms, both containing confluent Heun special functions [*11*], HeunC, with the same arguments, although the domains of the variables differ, as specified above. The complete solution of Schroedinger's partial-differential equation for the amplitude function as a product of solutions of the three ordinary-differential equations follows.

$$\psi = c\, N\, (\xi^2 - 1)^{\left(\frac{m}{2}\right)} (\eta^2 - 1)^{\left(\frac{m}{2} + \frac{1}{2}\right)} e^{\left(\frac{\pi d \sqrt{-2 E \mu}\,(\xi + \eta + 1)}{h} + i m \phi\right)} \mathrm{HeunC}\left(\frac{4\pi d \sqrt{-2 E \mu}}{h}, m,\right.$$
$$\left. m, \frac{2 Z e^2 \pi \mu d}{h^2 \varepsilon_0}, \frac{2 E \pi^2 \mu d^2}{h^2} + \frac{m^2}{2} + \_C1 - \frac{d Z e^2 \pi \mu}{h^2 \varepsilon_0}, \frac{\xi}{2} + \frac{1}{2}\right) \mathrm{HeunC}\left(\right.$$
$$\left. \frac{4\pi d \sqrt{-2 E \mu}}{h}, m, m, \frac{2 Z e^2 \pi \mu d}{h^2 \varepsilon_0}, \frac{2 E \pi^2 \mu d^2}{h^2} + \frac{m^2}{2} + \_C1 - \frac{d Z e^2 \pi \mu}{h^2 \varepsilon_0}, \frac{\eta}{2} + \frac{1}{2}\right) \Big/$$
$$\sqrt{2\pi}$$

Because the solutions of Heun's confluent differential equation [*11*] are more complicated, in that they admit no direct general expression in polynomial form, than those of Laguerre's or Legendre's equation that possesses fewer singularities, symbolic calculations involving the confluent Heun functions are difficult, precluding at present a direct derivation of energy *E* in terms of quantum numbers. On combining the exponential functions to form the total amplitude function, the coefficient of ½ *d* (ξ +η +1) must, on dimensional grounds and by comparison with the solutions in spherical polar and paraboloidal coordinates, be equal to (*n* $a_0$)$^{-1}$ in which $a_0$ denotes the Bohr radius,

$$a_0 = \frac{\varepsilon_0 h^2}{\pi m_e e^2}$$

and *n* is an energy quantum number from experiment, as in preceding solutions. In this manner we accept an expression for the energy, taking µ ≈ $m_e$,

$$E = -\frac{\mu Z^2 e^4}{8 h^2 \varepsilon_0^2 n^2} = -\frac{Z^2 h^2}{8 \mu a_0^2 \pi^2 n^2}$$

equivalent to that in preceding solutions. The total amplitude function independent of time contains only variables ξ, η, φ, quantum numbers *n* and *m*, atomic number *Z* and Bohr radius $a_0$, normalizing factor *N*, parameter *d* denoting the distance between the foci of the ellipsoids and parameter _C1 from the separation of Schroedinger's partial-differential equation into three ordinary-differential equations; the latter quantity, which appears in either confluent Heun function in an equivalent manner, becomes related to a third quantum parameter, λ. The eventual form of the amplitude function has this form.

$$\psi = N\sqrt{2}\,(\xi^2-1)^{\left(\frac{|m|}{2}\right)}(-\eta^2+1)^{\left(\frac{|m|}{2}+\frac{1}{2}\right)}e^{\left(m\phi I - \frac{dZ(\eta+\xi+1)}{2na_0}\right)}$$

$$\mathrm{HeunC}\left(\frac{2\,dZ}{n\,a_0},|m|,|m|,\frac{2Zd}{a_0},\frac{m^2}{2}-\frac{Zd}{a_0}+\lambda-\frac{Z^2d^2}{4n^2a_0^{\,2}},\frac{\xi}{2}+\frac{1}{2}\right)$$

$$\mathrm{HeunC}\left(\frac{2\,dZ}{n\,a_0},|m|,|m|,\frac{2Zd}{a_0},\frac{m^2}{2}-\frac{Zd}{a_0}+\lambda-\frac{Z^2d^2}{4n^2a_0^{\,2}},\frac{\eta}{2}+\frac{1}{2}\right)\Big/(2\sqrt{\pi})$$

When made discrete, parameter $\lambda$, which might have a symbolic expression in terms of distance $d$, is unlikely to assume integer values; the first argument of the HeunC functions must take positive values, but clearly cannot be integer, unlike the second and third arguments. The fourth argument also adopts a general value, positive or negative. The amplitude function in this form is the most concrete or explicit that is envisaged at present, pending further development of confluent Heun functions.

## V.     HYDROGEN ATOM IN SPHEROCONICAL COORDINATES

In this system of coordinates of which the variables have surfaces comprising a sphere and two double cones of necessarily elliptical cross section and along axes $x$ and $z$, a positively charged atomic nucleus lies near the origin, as in preceding cases, with distance $r$ to the reduced mass. The relations of spheroconical coordinates $\xi, \rho, \eta$ ($\xi$ and $\eta$ are distinct from those variables denoted with the same letters in ellipsoidal coordinates) with cartesian coordinates have this general form,

$$x=\frac{r\sqrt{(b^2+\xi^2)(b^2-\eta^2)}}{b},\ y=\frac{r\,\xi\,\eta}{a\,b},\ z=\frac{r\sqrt{(a^2-\xi^2)(a^2+\eta^2)}}{a}$$

but we apply a particular form with $a=\frac{1}{\sqrt{2}}, b=\frac{1}{\sqrt{2}}$ such that $a^2+b^2=1$. The jacobian for volume integrals is $\dfrac{4r^2(\eta^2+\xi^2)}{\sqrt{(1-4\xi^4)(1-4\eta^4)}}$. According to these conditions, Schroedinger's temporally independent partial-differential equation becomes, in SI units,

$$-h^2\left(-\frac{\xi^4\left(\frac{\partial^2}{\partial\xi^2}\psi(\xi,r,\eta)\right)}{r^2(\eta^2+\xi^2)}+\frac{\eta^2\left(\frac{\partial^2}{\partial r^2}\psi(\xi,r,\eta)\right)}{\eta^2+\xi^2}+\frac{\xi^2\left(\frac{\partial^2}{\partial r^2}\psi(\xi,r,\eta)\right)}{\eta^2+\xi^2}\right.$$

$$-\frac{\eta^4\left(\frac{\partial^2}{\partial\eta^2}\psi(\xi,r,\eta)\right)}{r^2(\eta^2+\xi^2)}-\frac{2\left(\frac{\partial}{\partial\xi}\psi(\xi,r,\eta)\right)\xi^3}{r^2(\eta^2+\xi^2)}+\frac{2\left(\frac{\partial}{\partial r}\psi(\xi,r,\eta)\right)\eta^2}{r(\eta^2+\xi^2)}$$

$$\left.+\frac{2\left(\frac{\partial}{\partial r}\psi(\xi,r,\eta)\right)\xi^2}{r(\eta^2+\xi^2)}-\frac{2\left(\frac{\partial}{\partial\eta}\psi(\xi,r,\eta)\right)\eta^3}{r^2(\eta^2+\xi^2)}+\frac{\frac{\partial^2}{\partial\xi^2}\psi(\xi,r,\eta)}{4r^2(\xi^2+\eta^2)}+\frac{\frac{\partial^2}{\partial\eta^2}\psi(\xi,r,\eta)}{4r^2(\xi^2+\eta^2)}\right)$$

$$\Big/(8\pi^2\mu)-\frac{Ze^2\psi(\xi,r,\eta)}{4\pi\varepsilon_0 r}=E\,\psi(\xi,r,\eta)$$

separable into these three ordinary-differential equations,

$$\frac{d^2}{d\xi^2}\Xi(\xi) = \frac{4\,\Xi(\xi)\,\xi^2\,c_2}{h^2\,\varepsilon_0\,(4\,\xi^4-1)} - \frac{\Xi(\xi)\,c_3}{h^2\,\varepsilon_0\,(4\,\xi^4-1)} - \frac{8\left(\frac{d}{d\xi}\Xi(\xi)\right)\xi^3}{4\,\xi^4-1},$$

$$\frac{d^2}{dr^2}R(r) = \frac{R(r)\,c_2}{r^2\,h^2\,\varepsilon_0} - \frac{2\left(h^2\left(\frac{d}{dr}R(r)\right)\varepsilon_0 + Z\,e^2\,R(r)\,\pi\,\mu + 4\,E\,R(r)\,\pi^2\,\mu\,r\,\varepsilon_0\right)}{h^2\,r\,\varepsilon_0}$$

and

$$\frac{d^2}{d\eta^2}H(\eta) = \frac{4\,H(\eta)\,\eta^2\,c_2}{h^2\,\varepsilon_0\,(4\,\eta^4-1)} + \frac{H(\eta)\,c_3}{h^2\,\varepsilon_0\,(4\,\eta^4-1)} - \frac{8\left(\frac{d}{d\eta}H(\eta)\right)\eta^3}{4\,\eta^4-1}$$

The normalized solution of this ordinary-differential equation for R(*r*),

$$R(r) = \sqrt{\frac{Z\,\pi\,\mu\,e^2\,k!}{\varepsilon_0\,h^2\,(1+2\,l+k)!}}\; e^{\left(-\frac{\pi\,\mu\,Z\,e^2\,r}{h^2\,\varepsilon_0\,(k+l+1)}\right)}\left(\frac{2\,\pi\,\mu\,Z\,e^2}{h^2\,\varepsilon_0\,(k+l+1)}\right)^{(l+1)}\,r^l$$
$$\mathrm{LaguerreL}\left(k,\,2\,l+1,\,\frac{2\,\pi\,\mu\,Z\,e^2\,r}{h^2\,\varepsilon_0\,(k+l+1)}\right)/(k+l+1),$$

is identical to the corresponding solution in spherical polar coordinates. The other two physically acceptable solutions, which contain general Heun functions [*11*] present as HeunG,

$$\Xi(\xi) = N_\xi\sqrt{1-2\,\xi^2}\;\mathrm{HeunG}\left(-1,\,\kappa+\frac{1}{4},\,1+\frac{l}{2},\,-\frac{l}{2}+\frac{1}{2},\,\frac{1}{2},\,\frac{1}{2},\,-2\,\xi^2\right)$$

and

$$H(\eta) = N_\eta\sqrt{1-2\,\eta^2}\;\mathrm{HeunG}\left(-1,\,-\kappa+\frac{1}{4},\,1+\frac{l}{2},\,-\frac{l}{2}+\frac{1}{2},\,\frac{1}{2},\,\frac{1}{2},\,-2\,\eta^2\right),$$

have identical forms except that the signs before parameter κ, in the second argument of each Heun function, differ; of the two independent solutions, both containing these Heun functions, of each ordinary-differential equation for $\Xi(\xi)$ and $H(\eta)$, the solutions other than those above are physically unacceptable. The total amplitude function is hence

$$\psi(\xi,\,r,\,\eta) = N\,\Xi(\xi)\,R(r)\,H(\eta)$$

in which separate normalizing factors $N_\xi$ and $N_\eta$ become merged into a single factor, denoted *N*; to simplify the form, we replace various fundamental constants with Bohr radius $a_0$.

$$\psi(\xi, r, \eta) = c\, N \sqrt{\frac{Z\, k!}{(1+2l+k)!\, a_0}} \left(\frac{2Z}{a_0(k+l+1)}\right)^{(l+1)} r^l\, e^{\left(-\frac{Z\, r}{a_0(k+l+1)}\right)}$$

$$\mathrm{LaguerreL}\!\left(k, 2l+1, \frac{2Z\, r}{a_0(k+l+1)}\right) \sqrt{1 - 2\xi^2}$$

$$\mathrm{HeunG}\!\left(-1, \kappa + \frac{1}{4}, 1 + \frac{l}{2}, -\frac{l}{2} + \frac{1}{2}, \frac{1}{2}, \frac{1}{2}, -2\xi^2\right) \sqrt{1 - 2\eta^2}$$

$$\mathrm{HeunG}\!\left(-1, -\kappa + \frac{1}{4}, 1 + \frac{l}{2}, -\frac{l}{2} + \frac{1}{2}, \frac{1}{2}, \frac{1}{2}, -2\eta^2\right) / (k+l+1)$$

A notable property of this solution is that, with coefficient $c$ arbitrarily set equal to +1, it is entirely real, i.e. has no imaginary parts, which in the three preceding systems of coordinates resulted from at least the presence of an exponential term involving angular coordinate $\phi$. For the same reason, this solution lacks equatorial quantum number $m$, with which is associated the loss of degeneracy of states with common energy quantum number $n$ on application of an external magnetic field to a H atom. This absence of $m$ in no way implies the retention of degeneracy in the presence of a magnetic field; an explicit calculation including the effect of an homogeneous magnetic field is just as applicable in spheroconical coordinates as in spherical polar coordinates. According to the coefficient of $t$ in the temporal exponent (not shown), or the coefficient of $r$ in the included exponential term, the energy of a discrete state associated with an amplitude function characterized with quantum numbers $k$, $l$, $\kappa$ depends on only quantum numbers $k$ and $l$ in combination, so is independent of $\kappa$, as is confirmed with direct calculations of $E = \int \psi\, H \psi\, dv$ with hamiltonian $H$, normalised amplitude function $\psi$ and many and varied values of $k$, $l$, $\kappa$; we hence relate the experimental quantum number $n$ for energy according to the same theoretical relation as for spherical polar coordinates, so $n = k + l + 1$.

## VI. APPLICATIONS OF AMPLITUDE FUNCTIONS

According to wave mechanics, any observable property of a H atom is calculable with amplitude functions expressed in the appropriate coordinates when the respective operator is included in a suitable form. Two important properties of a state of an H atom are the square of the total electronic angular momentum, $L^2$, excluding the intrinsic angular momentum (spin) of electron or proton, and the component of total angular momentum, $L_z$, parallel to a particular coordinate polar axis, chosen to be cartesian axis $z$. $\psi_{k,l,m}(r,\theta,\phi)$, $\psi_{n1,n2,m}(u,v,\phi)$ and $\psi_{n,m,\lambda}(\xi,\eta,\phi)$ are all eigenfunctions of operator $L_z$ for that component; the common eigenvalue is

$$L_z\, \psi = m\, h\, \psi / 2\pi,$$

in terms of equatorial quantum number $m$, but only spherical polar $\psi_{k,l,m}(r,\theta,\phi)$, and spheroconical $\psi_{k,l,\kappa}(\xi,r,\eta)$ are eigenfunctions of operator $L^2$, as

$$L^2\, \psi_{k,l,m}(r,\theta,\phi) = l(l+1)\, h^2\, \psi_{k,l,m}(r,\theta,\phi) / 4\pi^2,$$
$$L^2\, \psi_{k,l,\kappa}(\xi,r,\eta) = l(l+1)\, h^2\, \psi_{k,l,\kappa}(\xi,r,\eta) / 4\pi^2,$$

with the eigenvalue as an expression containing azimuthal quantum number $l$. For paraboloidal coordinates, we derive instead an expectation value of $L^2$ for any amplitude function defined with quantum numbers $n_1$, $n_2$, $m$ as

$$<L^2> = \frac{1}{4} \frac{((n_1 + n_2 + 1)|m| + m^2 + n_1 + 2 n_1 n_2 + n_2) h^2}{\pi^2}$$

which accordingly depends on all those three quantum numbers in combination, like the energy of that state. Although from that formula one might derive an expression for $l$,

$$l = -\frac{1}{2} + \sqrt{(|m| + n_1 + n_2 + 1)(|m| + 1) + 2 n_1 n_2 - |m| - \frac{3}{4}}$$

which clearly assumes no integer or half-integer value except when $n_1 = n_2 = 0$ giving $l = |m|$; taking $|m|$ as being a lower limit of $l$ is clearly a preferable interpretation.

A quantity related to operators for angular momentum is Runge-Lenz-Pauli operator $A$, which one might consider to resemble a fourth component of angular momentum arising from a symmetry that the hydrogen atom shares with rotation group $O(4)$ in four dimensions [7]. For a stationary state associated with an amplitude function defined with three pertinent quantum numbers, $\frac{\partial}{\partial t} E = 0$, $\frac{\partial}{\partial t} L = 0$ and $\frac{\partial}{\partial t} A = 0$, as all these quantities are conserved. In spherical polar coordinates, amplitude functions $\psi(r,\theta,\phi)$ are not eigenfunctions of operator $A$; its component $A_z$ generates two functions corresponding to the same energy but with distinct values of quantum numbers $k$ and $l$,

$$A_z \psi_{k,l,m} = -\frac{\sqrt{(k+l+1)^2 - (l+1)^2} \sqrt{\frac{(l+m+1)(l-m+1)}{(2l+1)(3+2l)}} \psi_{k-1, l+1, m}}{k+l+1}$$
$$- \frac{\sqrt{(k+l+1)^2 - l^2} \sqrt{\frac{(l+m)(l-m)}{(2l+1)(2l-1)}} \psi_{k+1, l-1, m}}{k+l+1}$$

in which $k$ and $l$ alter in opposite senses but $n = k + l + 1$ remains constant. Like operator $L$ for angular momentum, for which amplitude function $\psi_{k,l,m}$ is not an eigenfunction whereas this property holds for operator $L^2$, the eigenvalues of $A^2$ conform to this formula.

$$A^2 \psi_{k, l, m} = \left(1 - \frac{Z^2 (l^2 + l + 1)}{(k+l+1)^2}\right) \psi_{k, l, m}$$

In contrast, in paraboloidal coordinates, $\psi_{n_1, n_2, m}$ is an eigenfunction of $A_z$, according to this formula,

$$A_z \psi_{n_1, n_2, m} = \frac{(n_2 - n_1) Z}{n_1 + n_2 + |m| + 1} \psi_{n_1, n_2, m}$$

in which difference $n_2 - n_1$, or its reverse, might be called an *electric quantum number*, because the energy shift of the linear Stark effect, whereby an H atom interacts with an external electric field, depends on that difference [14]. For spherical polar coordinates, the three quantities that yield eigenvalues are thus the energy, through the hamiltonian operator, the square of the total angular momentum, $L^2$, and its component $L_z$, for which the eigenvalues are specified above. For paraboloidal coordinates, the three respective quantities are energy and components $L_z$ and $A_z$, the latter hence replacing $L^2$. For ellipsoidal coordinates, the three quantities are the energy, $L_z$ and another parameter arising from the separation of variables, whereas for spheroconical coordinates the three quantities are energy, the square of total angular momentum $L^2$ and another parameter also arising from the separation of variables [15].

An observable property of a H atom is its absorption spectrum, in terms of both the frequencies and the intensities of the discrete spectral lines associated with transitions between states for which the discrete energies have $E < 0$. Expressed as an oscillator strength or $f$ value, as defined above, the intensity of a spectral line is proportional to the square of an electric dipolar moment for a transition, calculated in paraboloidal coordinates as a matrix element of $\frac{1}{2} e (u^2 - v^2)$ between amplitude functions associated with two combining states. For this purpose we make an association, in Dirac's bracket notation, between amplitude functions defined with spherical polar and paraboloidal quantum numbers,

$$|k=0, l=1, m=0> = \frac{\sqrt{2}}{2} (|n_1=0, n_2=1, m=0> - |n_1=1, n_2=0, m=0>)$$

with analogous appropriate combinations as sums for amplitude functions with $k > 0$. For the particular state involving $|0,0,0>$ and the right side above, the result of an integration involving the amplitude functions in paraboloidal coordinates depending on spatial coordinates is

$$<e\,z> = \frac{128\sqrt{2}\,\varepsilon_0\,h^2}{243\,\pi\,Z\,e\,\mu}$$

This formula is necessarily exactly the same as that in spherical polar coordinates, apart from the temporal factor in the latter, derived above. Despite the disparity between the effective size of the H atom, ~ 0.1 nm, and wave length 121.6 nm of light for this transition, we interpret, classically, the electric vector of the electromagnetic wave to induce an electric-dipolar moment for the transition in the atom that oscillates at the same frequency, such that in absorption radiant energy is transferred to become internal electronic energy of that atom.

With this scheme and the formula for $f$ in terms of the matrix element for dipolar moment involving amplitude functions with quantum numbers $n_1$, $n_2$ in appropriate combinations,

$$f = \frac{8\pi^2 m_e \nu}{h\,e^2}\,|<n_1, n_2, 0 | e\left(\frac{u^2}{2} - \frac{v^2}{2}\right)|0,0,0>|^2$$

containing frequency $\nu$ as the coefficient of $it$ in the temporal factor (not shown), we calculate the frequencies and intensities of the first ten lines in the Lyman series in absorption in the vacuum-ultraviolet region. Derived directly from these paraboloidal amplitude functions, the resulting spectrum appears in figure 1; this spectrum is identical with that calculated with amplitude functions in spherical polar coordinates [10], which proves the equivalence of the calculations with coordinates in disparate sets and with respective quantum numbers in disparate sets. The continuous spectrum for energies greater than the threshold for ionization is generated analogously with amplitude functions in paraboloidal coordinates just as for spherical polar coordinates [*10*], and is calculated likewise in ellipsoidal and spheroconical coordinates, apart from increased complication because of the Heun functions in the pertinent amplitude functions.

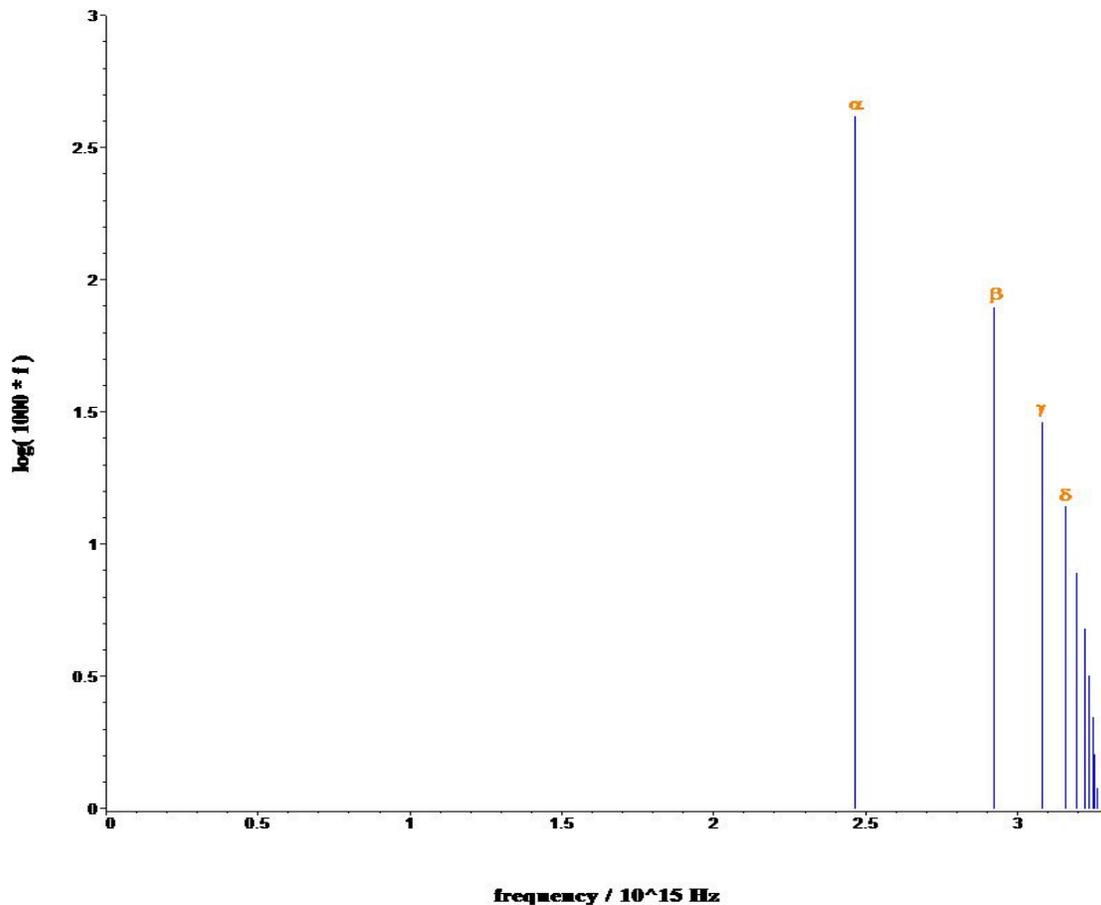

**FIGURE 1.** Absorption spectrum for discrete transitions of the H atom in the vacuum-ultraviolet region quantitatively calculated with amplitude functions in *paraboloidal* coordinates; the intensity is plotted as $\log_{10}(1000\,f)$ against frequency/$10^{15}$ Hz of the transitions. The first four lines of the Lyman series are identified with greek letters.

## VII.  DISCUSSION

For a free particle and a canonical linear harmonic oscillator in multiple dimensions and for a hydrogen atom, Schroedinger's partial-differential equation independent of time, like Laplace's equation of which the laplacian operator is included in the former, is *super-integrable* -- it admits separable and integrable solutions in multiple coordinate systems.  For the hydrogen atom, the systems of coordinates were delineated to number precisely four [*7*]; a subsequent derivation of these systems proved their correctness [*15*].  A claim of a fifth system of coordinates [*16*] is misleading because the alleged *rectangular* coordinates still contain radial distance *r* in set (*x,y,z,r*); the spherical harmonics in the general spherical polar coordinates become replaced therein with functions involving ratios of *x*, *y*, *z* [*16*]. Advanced mathematical software, here *Maple*, facilitates quantitative and exact symbolic calculations in molecular and chemical physics; the particular capability to solve directly a partial-differential equation enables a user to avoid the mathematical contortions of a traditional manual derivation [*17*] of amplitude functions of an atom with one electron treated according to Schroedinger's formulation of wave mechanics involving partial-differential equations of second order, instead to focus on the properties of the solutions. A user can still test the solutions to prove their correctness, and transform them as desired; for instance, the direct solution of the ordinary-differential radial equation as part of Schroedinger's equation in either spherical polar or paraboloidal coordinates appears in terms of Whittaker functions, but a transformation, either directly or indirectly through other functions, into further functions in the same coordinate system, such as Laguerre functions that appear in traditional derivations, is an option readily implemented. Here we apply this capability to solve Schroedinger's equation, temporally dependent or independent, in four systems of coordinates -- spherical polar, paraboloidal, ellipsoidal and spheroconical, simply on generating an appropriate laplacian operator and specifying correctly the electrostatic potential energy within the hamiltonian operator according to the particular coordinates, then commanding the solution of the resulting partial-differential equation. Our argument in deriving a solution in paraboloidal coordinates naturally yields the same result as that which appears in some physics textbooks; one source [*14*] includes a normalizing factor of these functions that is incorrect, and in other books this factor is not specified explicitly. The application of boundary conditions to evaluate the separation parameters invokes a further algebraic capability that is formally directly available; the somewhat intractable confluent Heun functions [*11*] hinder at present this application for ellipsoidal parameter $\lambda$. Calculations with general Heun functions [*11*] are easier than with confluent Heun functions; we calculated numerical normalizing factors and energies of $\psi(\xi,r,\eta)$ for many and diverse values of quantum numbers $k$, $l$ and $\kappa$. The graphical capability of the same software is then brought to bear to generate accurate plots of pertinent properties, such as the absorption spectrum of H presented quantitatively in figure 1 in two dimensions and the surfaces depicting amplitude functions at particular values of $\psi$ in other papers in a subsequent series.

    A choice of amplitude functions in the four systems of coordinates must depend on the purpose of a calculation on the hydrogen atom, or other atom with only one electron, in which the amplitude functions serve as working formulae according to wave mechanics. The overwhelmingly best known system has, of course, spherical polar coordinates; as the properties of the Laguerre and Legendre polynomials [*11*] involved therein are highly developed, calculations are generally rapid. This system is applicable to a hydrogen atom, or to any other atom with only one electron, that is in isolation -- no other matter in the vicinity, no applied electric field apart from an electromagnetic wave in the form of light that might interact classically with the atom in absorption, emission or scattering. A uniform external magnetic field causes a loss of the degeneracy associated with equatorial quantum number $m$. The authors of textbooks on quantum mechanics in physics typically content themselves with the mathematical details of this solution of the temporally independent Schroedinger equation and present some exemplary formulae for a few selected functions. Practically all textbooks of chemistry allude to these functions in spherical polar coordinates, generally in mistaken contexts; some such textbooks, particularly in physical and inorganic chemistry, describe their properties with accurate formulae but more or less inaccurate figures depicting poorly defined surfaces and shapes. Following Schroedinger's own solution of his equation in paraboloidal coordinates [*6*], some textbooks of quantum mechanics in physics treat this system, but no known textbook of chemistry even mentions that this system exists for the hydrogen atom. Common to spherical polar and paraboloidal amplitude functions, Laguerre polynomials [*11*], for two spatial variables in the latter system, are readily manipulated, and calculations are generally rapid. Schroedinger applied this system to treat, with perturbation theory that he concurrently developed, the hydrogen atom in an homogeneous electric field; the purpose was to calculate the Stark effect, explicitly the shifting, splitting and intensities of spectral

lines as a result of hydrogen atoms being subjected to a uniform electric field. Other contexts in which these paraboloidal coordinates are directly useful include the photoelectric effect, the Compton effect and a collision of an electron with a H atom [14]; in each case a particular direction in space is distinguished according to some external force. In ellipsoidal coordinates, one focus of an ellipsoid is located at or near the atomic nucleus, and another focus, at distance $d$, is merely a dummy location; as the latter can become the location of a second atomic nucleus, the associated amplitude functions become formally applicable to a diatomic molecule, which has been the reason for the modest attention devoted to these coordinates. These amplitude functions as directly derived here, for the first time, contain confluent Heun functions, which pose difficulties of calculation because they lack a simple polynomial expression [11]. Some indirect derivations of amplitude functions in the literature, through solutions of the differential equations in series, have been implemented for that reason [8]; the shapes of surfaces of these functions at a particular value of $\psi$ depend strongly on that distance $d$. For all three preceding coordinate systems, equatorial angle $\phi$ is one variable; its presence in a resulting derived amplitude function has invariably the form stated above, in which the presence of $i = \sqrt{-1}$ as factor of equatorial quantum number $m$ in the exponent dictates mostly complex total amplitude functions, even with coefficient $c = 1$; their intrinsic real and imaginary parts hence preclude depiction of most surfaces directly in real space, even with a favoured value of external coefficient $c$. As a further complication, confluent Heun functions in ellipsoidal coordinate $\xi$ have generally a complex nature. In contrast, with $c = +1$ or $-1$, each and every amplitude function in spheroconical coordinates is entirely *real* -- thus has no imaginary part, enabling a direct plot of each such surface. Calculations with the general Heun functions in two spheroconical coordinates $\xi$ and $\eta$ are easier than with confluent Heun functions in their two coordinates; the third spheroconical coordinate is just the separation $r$ between the reduced mass and the origin, essentially between the electron and nucleus, the same as in spherical polar coordinates. These coordinates have thus much to recommend them for a general discussion of the wave-mechanical properties of the hydrogen atom. For realistic and practical purposes, the solutions in paraboloidal and ellipsoidal coordinates have, in contrast, significant applications, as mentioned above, whereas the solutions in either spherical polar or spheroconical coordinates serve merely as exercises in the solutions of partial-differential equations.

## VIII. CONCLUSION

Applying powerful mathematical software, we undertook direct calculations in wave mechanics, according to Schroedinger's formulation, for the hydrogen atom in coordinates in four systems, two more than by Schroedinger himself. The results indicate that the observable properties of H, such as the angular momenta of particular states and the absorption spectrum associated with transitions between discrete states, are directly calculable in any of these four coordinate systems. For a particular purpose, one system of coordinates might prove more convenient than another; for instance, for the calculation of the effect of an externally applied uniform electric field on the H atom, known as the Stark effect, paraboloidal coordinates are preferable to spherical polar coordinates, as originally applied by Schroedinger [6], but that calculation is equally practicable in other coordinates. The advantage of ellipsoidal coordinates arises in calculations on diatomic-molecular species, in which a separate atomic nucleus might be located at the second focus of the ellipsoids. Two advantages of amplitude functions in spheroconical coordinates are that they can be directly taken to be entirely real, hence having no imaginary parts, and that with each amplitude function are associated two quantum numbers $k$ and $l$ that define in combination the energy and $l$ separately the angular momentum, but these solutions are applicable purely to an isolated hydrogen atom that again makes them of negligible chemical and physical interest. The disadvantage, at present, of both ellipsoidal and spheroconical coordinates is that they each involve Heun functions [*11*], of which applications are less well developed than for Laguerre functions that arise in both spherical polar and paraboloidal coordinates. Distinct from the amplitude functions, the electronic states of a H atom are specified rigorously within the Schroedinger formalism in terms of energy quantum number $n$ and azimuthal quantum number $l$; the latter quantum number arises naturally in solutions in both spherical polar and spheroconical coordinates, and indirectly in paraboloidal and ellipsoidal coordinates as $|m| \leq l$. *The shape of an amplitude function, and even the quantum numbers in a particular set to specify such an individual function, depend on the coordinates in a particular chosen system, and are therefore artefacts of a particular coordinate representation within wave mechanics*, which includes both coordinate and momentum representations and which is just one method among many in quantum mechanics [*3*,*4*].

An amplitude function of any kind discussed here conforms to a definition of *orbital* as *one-electron wave function such as the hydrogen-like wave functions* [*19*]; such an orbital has no tangible existence [*20*], and with its particular quantum numbers is an artefact not only of a particular quantum-mechanical method of calculation, namely wave mechanics, but also of a particular representation – either coordinate or momentum, and, furthermore, a particular variable set therein. Each coordinate system, or a momentum representation in its respective guises, would have equally parochial orbitals and attributes, but the energy and angular momenta of an atom in a particular state are immutable; specific values of angular momenta might be associated with amplitude functions in various combinations with common $n$. A transformation of coordinates enables one to express an amplitude function derived directly in any system of coordinates in terms of amplitude functions in linear combinations belonging to the same value of energy, hence $n$, in any other system. For instance, the graphical displays of surfaces of spheroconical $\psi_{1,0,0}(\xi,r,\eta)$ and spherical polar $\psi_{1,0,0}(r,\theta,\phi)$ at appropriate values of $\psi$ demonstrate directly the correlation between these two amplitude functions. The surface of $\psi_{1,0,0}(u,v,\phi)$ in paraboloidal coordinates is identical with that of a *sp* or diagonal hybrid orbital in spherical polar coordinates, which is simply a linear combination of $\psi_{0,1,0}(r,\theta,\phi)$ and $\psi_{1,0,0}(r,\theta,\phi)$. A transformation involving ellipsoidal coordinates must take into account distance $d$ between the foci of the ellipsoid. The definition of an orbital [*19*] as a mathematical formula, specifically that of a one-electron function as a solution of Schroedinger's temporally independent equation, is applicable to algebraic formulae having as variables not only spatial coordinates but also components of momenta, because one can choose between a coordinate representation and a momentum representation as a basis of solution of that equation. For each solution in coordinates of a particular system, there is a corresponding solution in momentum components. The simplistic conversion from spherical polar coordinates to the polar momenta by Podolski and Pauling [*23*] was erroneous because those momenta components were not orthogonal, as Lombardi [*24*] noted with a correct transformation. Klein [*25*] undertook a transformation of amplitude functions from paraboloidal coordinates to toroidal variables for components of momenta, but the orthogonality of those variables was not tested. Other specifications of the amplitude functions in the coordinate representations exist [for instance *7*, *26*] but these are merely formal solutions not directly amenable to practical calculations such as the plots of surfaces and derivation of other properties. There hence exist at least ten sets of formulae for orbitals [*7*], each with its set of quantum numbers and distinct shapes of the respective surfaces.

A failure to distinguish between a state of an atom and a postulated amplitude function, in whatever coordinates, constitutes a *category fallacy*. A state of an atom is defined by its energy and its angular momentum; an amplitude function is defined by not only its quantum parameters but also its coordinate system. Quantum mechanics in its pioneer forms -- matrix mechanics and Schroedinger's wave mechanics, both non-relativistic -- must yield the same values of observable quantities for a particular state or transition in whatever system of coordinates. In matrix mechanics [21] there is no amplitude function; instead of applying $H \psi = E \psi$, or its integral form, in wave mechanics with $H$ as an hamiltonian operator pertaining to a particular coordinate or momentum representation and involving differential quantities appropriate to that representation, one forms an appropriate hamiltonian matrix **H** that is made diagonal to produce energy matrix **E**. Matrix mechanics, unlike wave mechanics, is thus applicable to calculations involving intrinsic angular momenta pertinent to experiments in magnetic resonance, for instance the calculation of the spectrum of a 'spin' system given the values of chemical shifts and coupling parameters. Another approach [22] that has its basis in the symbolic calculations of Pauli on the hydrogen atom [5], undertaken before Schroedinger's publications [6], that is entirely symbolic might also be applicable to an extensive treatment of a hydrogen atom without such artefactual properties as amplitude functions parochial to a particular system of coordinates. As a state of atomic H is defined uniquely according to its energy and its angular momenta, any other attribute of that state that is an artefact of one or other representation must be recognized as such. For only an atom with one electron does an energy of a discrete state, relative to a reference state such as a threshold of ionization, identify uniquely with a specific value of a particular quantum number or quantum parameter.

**Acknowledgement**

I am indebted to many colleagues for helpful comments.